# Time Evolution of the Microwave Second-Harmonic Response of MgB$_2$ Superconductor

A. Agliolo Gallitto[*], M. Li Vigni and G. Vaglica

*INFM and Dipartimento di Scienze Fisiche e Astronomiche, Università di Palermo, via Archirafi 36, I-90123 Palermo (Italy)*

**Abstract**

We report on transient effects in the microwave second-order response of two ceramic MgB$_2$ samples. The time evolution of the second-harmonic signal is investigated for about 500 s after the sample has been exposed to a variation of the dc magnetic field. We suggest that during the first seconds the response is determined by diffusive motion of fluxons, while in the time scale of minutes it is ruled by magnetic relaxation over the surface barrier.

*Keyword*s: Fluxon dynamics; magnetic relaxation; nonlinear microwave response.
*Pacs*. 74.25.Ha;74.25.Nf; 74.60.Ge

## 1. Introduction

It is well known that the presence of the surface barrier strongly affects the irreversible proprieties of superconductors. Interaction between magnetic flux and sample surface has mainly been studied by magnetization measurements [1-4]. Surface-barrier effects manifest themselves in: i) first-penetration field, $H_p$, higher than the lower critical field, $H_{c1}$; ii) hysteresis loop of the magnetization curve asymmetric in the two branches at increasing and decreasing fields; iii) magnetic relaxation rates different for flux entry and exit.

Recently, we have detected the relaxation of the microwave (mw) second-order response of superconductors in the mixed state. We have observed that, after the dc magnetic field has reached a certain value, in the time scale of minutes the signal radiated at the second-harmonic (SH) frequency of the driving field decays following a logarithmic law. Further, the relaxation rate of the SH signal depends on the way in which the dc magnetic field is reached, i.e. at increasing or decreasing values. This finding has suggested that the decay in that time scale is determined by fluxon motion through the surface barrier [5].

In this paper we investigate the time evolution of the SH signal in two MgB$_2$ samples, a bulk sample and a powdered one. We show that, after the sample has been exposed to a variation of the dc magnetic field, the SH signal exhibits an initial exponential decay, which lasts for about 10 s, and a logarithmic decay, in the time scale of minutes. Our results suggest that magnetic relaxation over the surface barrier can be highlighted by measuring the time evolution of the mw SH response.

―――――
[*] Corresponding author. Tel.: +39-091-6234207; fax: +39-091-6162461; e-mail: agliolo@fisica.unipa.it.

## 2. Experimental

Time evolution of the SH signal has been studied in two different ceramic $MgB_2$ samples, with $T_c \approx 39$ K. The samples have been extracted from the same pellet, sintered from Alfa-Aesar powder at 800°C in Ar atmosphere, for three hours. One is a bulk sample of approximate dimensions $2 \times 1.5 \times 1$ mm$^3$; the other consists of $\approx 6$ mg of fine powder, obtained by grinding a piece of the ceramic material.

The sample is placed in a bimodal cavity, resonating at the two angular frequencies $\omega$ and $2\omega$, with $\omega/2\pi \approx 3$ GHz, in a region in which the mw magnetic fields $H(\omega)$ and $H(2\omega)$ are maximal and parallel to each other. The $\omega$-mode of the cavity is fed by a train of mw pulses, with pulse width 5 $\mu$s and pulse repetition rate 200 Hz. The maximal peak power is $\approx 50$ W. The harmonic signals radiated by the sample are detected by a superheterodyne receiver. The cavity is placed between the poles of an electromagnet, which generates dc magnetic fields, $H_0$, up to $\approx 10$ kOe. All measurements have been performed at $T = 4.2$ K with $H_0 \| H(\omega) \| H(2\omega)$.

Before any measurement was performed the sample was zero-field cooled down to $T = 4.2$ K; $H_0$ was increased up to 10 kOe and then decreased down to the residual field of the electromagnet. This preliminary procedure ensures that SH signals arising from processes occurring in weak links are suppressed by the trapped flux. The dc field was then swept with constant rate up to fixed values and the evolution of the SH signal was measured for $\sim 500$ s from the instant in which each $H_0$ value has been reached, with sampling time $\approx 0.3$ s.

Fig.1 shows the time evolution of the SH signal measured from the instant in which $H_0$ has reached the value of 3 kOe, on increasing (open symbols) and decreasing (full symbols) the field. In panel a) we report the results obtained in the powder. Measurements in the bulk sample have been performed for two different orientations of the sample with respect to $H_0$: the results obtained with $H_0$ perpendicular and parallel to the largest surface are reported in panels b) and c), respectively. It is worth remarking that the largest surface corresponds to the pristine surface of the pellet from which the sample was extracted; while the others have been obtained by cutting the pellet with a diamond saw. During the first $\sim 10$ s all the SH-vs-$t$ curves show similar behavior. In this time scale, the SH signal shows an exponential decay. However, at longer times the signal decay observed for the bulk sample is different in the two orientations. For $t > 10$ s, the SH signal shown in panel c) is stationary; on the contrary, those in panels a) and b) decay with a logarithmic law.

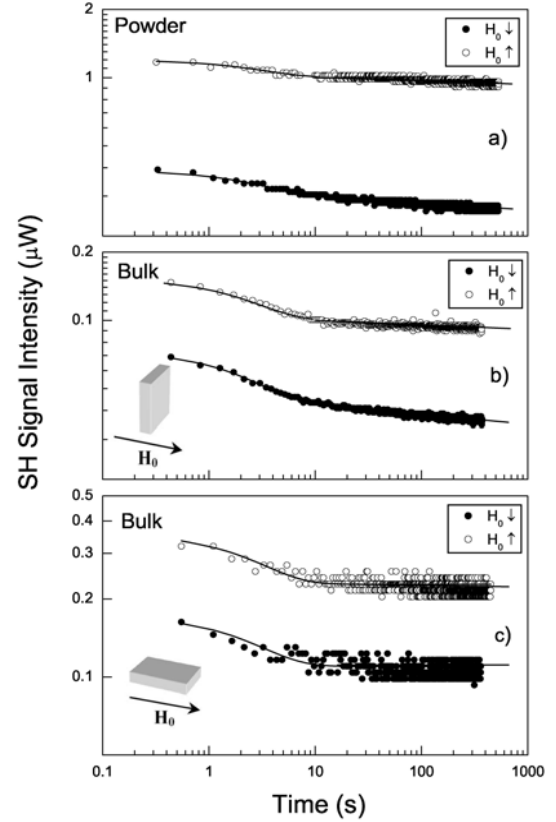

Fig.1. Time evolution of the SH signal at $H_0 = 3$ kOe. $T = 4.2$ K; input peak power $\approx 2$ W. Symbols are experimental data; lines are the best-fit curves obtained as explained in the text.

In order to deduce the parameters characteristic of the SH-signal decay, we have fitted the experimental data by the following expressions:

$$SH = A + B\, exp(-t/\tau) \qquad 0 < t < 10\ s \qquad (1)$$

$$SH = C\, [1 - D\, log(t/t_0)] \qquad t > 10\ s, \qquad (2)$$

with $t_0 = 10$ s and $A = C$ the value that the intensity of the SH signal takes on at $t = t_0$.

The lines in Fig. 1 are the best-fit curves. The values of the parameter $\tau$ fall in the range 3 s ± 20 %, independently of the sample, the geometry and the field-sweep direction. On the contrary, the value of the parameter $D$ depends on the way in which the field has been reached.

Measurements performed at different values of the dc magnetic field have shown that the characteristic time of the exponential decay is roughly independent of $H_0$, within the experimental accuracy.

Fig.2 shows the value of the best-fit parameter $D$ as a function of $H_0$. The rate of the logarithmic decay depends on $H_0$ and its sweep direction; it is larger for decreasing than for increasing fields. On increasing $H_0$, the logarithmic-decay rates for negative and positive field variations approach each other.

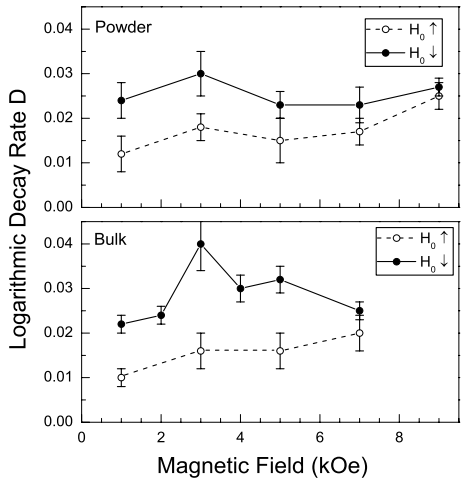

Fig.2. Field dependence of the logarithmic-decay rate, for the powdered and the bulk sample (with $H_0$ perpendicular to the largest surface). Open and full symbols refer to measurements performed at increasing and decreasing fields, respectively.

A further peculiarity of the SH response of both samples, which can be seen in Fig.1, is its hysteretic behavior; the SH signal is more intense at increasing than at decreasing fields. Measurements at different temperatures have shown that up to ≈ 35 K the hysteretic behavior is present in the whole range of magnetic fields investigated (0÷10 kOe); on further increasing the temperature, the field range in which the hysteresis is present shrinks. Furthermore, the SH signal decays only for the $H_0$ values at which the hysteresis is observed; when the hysteresis is missing the SH signal is stationary. Therefore, we infer that the hysteresis is strictly related to the transients.

## 3. Discussion

Microwave SH emission by superconductors in the critical state has been ascribed to a rectification process of the mw field operated by the superconductor [6]. The process arises because of the inertia of the fluxon lattice to follow high-frequency fields. It has been supposed that, due to the rigidity of the fluxon lattice, the induction field inside the sample does not follow adiabatically the mw field variations, except when the variations bring about motion of fluxons in the surface layers of the sample. From this model, it is expected that superconductors in the critical state radiate stationary SH signals, whose intensity is independent of the magnetic-field-sweep direction. Nevertheless, in the framework of the model, the time evolution of the SH signal can arise by supposing that the fluxon configuration near the sample surface evolves on elapsing the time.

At low temperatures, after the samples have been exposed to magnetic fields larger than the first penetration field, it is reasonable to hypothesize that a critical state develops; so, SH emission can be ascribed to the rectification process. The question is what mechanisms are responsible for the SH-signal decay.

Magnetic relaxation occurs through several mechanisms: diffusive motion of fluxons, surmounting of the surface and pinning barriers. Diffusive processes give rise to an exponential decay of the magnetization, while relaxation over bulk pinning gives rise to a logarithmic one. The effects of the fluxon motion over the surface barrier generally manifest with different decay rates for flux entry and exit; however, the ratio between the relaxation rates for flux entry and exit depends on the time window within which the relaxation is observed [7].

Our results point out that, during the first ~ 10 s after the sweep of the dc magnetic field has been stopped, the SH signal shows an exponential decay independently of the shape of the sample and its orientation with respect to $H_0$. Exponential decay of the dc magnetization occurring in the time scale of seconds has been reported in Ref. [8]; it has been discussed in the framework of the Brandt procedure [9], put forward for investigating diffusive motion of fluxons induced by variations of the applied field. The similarity of our results and those of Ref. [8] suggests that diffusive motion of fluxons is responsible for the initial exponential decay of the SH signal.

In the time scale of minutes, we have observed a logarithmic decay of the SH signal, characterized by rates of relaxation different for positive and negative field variations. According to the Burlachkov theory [7], the initial stage of the magnetic relaxation should be determined by the weakest one of the two sources of magnetic irreversibility: the bulk pinning and the surface barrier. It has been shown that the magnetic hysteresis loops of powdered $MgB_2$ samples show asymmetric behavior in the two branches at increasing and decreasing fields [4,10], suggesting that the surface barrier controls the magnetic irreversibility. The results have been ascribed to the weakness of the bulk pinning in the powder [4]; so, the logarithmic decay of the SH signal could be ascribed to flux-creep processes, which occur at short times because of the weak bulk pinning. However, several results disagree with this hypothesis. Firstly, relaxation by flux creep should be strongly affected by the temperature. We have measured the time evolution of the SH signal at different values of the temperature. The results of this investigation, not reported here, have shown that, for both samples, the peculiarities of the SH-signal decay are not significantly affected by the temperature. Furthermore, relaxation over bulk-pinning barrier does not justify the dependence of the relaxation rate on the way the magnetic field has been reached. On the other hand, comparison between the results of Fig.s 1 b) and c) show that the time evolution of the SH signal in the time scale of minutes strongly depends on the surface of the sample through which the magnetic field penetrates. Since for ceramic samples it is unreasonable supposing that the effects of the bulk pinning depend on the direction of the field penetration, we may ascribe the logarithmic decay of the SH signal of Fig. 1 b) to magnetic relaxation over the surface barrier. Furthermore, the similarity between the results of Fig.s 1 a) and b) suggests that even for the powdered sample the logarithmic decay of the SH signal is ascribable to the motion of fluxons over the surface barrier.

The finding that the decay rates of the SH signal are not sensitive to the temperature variation rules out the possibility that relaxation over the pinning barrier affects the SH emission. At present, the reason for that is not fully understood; probably, this is so because the mw harmonic emission is related to processes occurring in the surface layers of the superconducting sample.

In conclusion, we have reported on transient effects in the mw second-order response of $MgB_2$ superconductor. We have shown that during the first seconds after the magnetic field sweep has been stopped the SH signal decays exponentially, while in the time scale of minutes it shows a logarithmic decay. The time evolution in the time scale of minutes strongly depends on the sample surface, so it has been ascribed to surface-barrier effects. We suggest that the whole process responsible for the SH-signal decay can be figured out as follow. Following the Clem theory [11], when $H_0$ is varying the surface barrier is absent provided that $H_0 > H_{en}$, for positive field variations, or $H_0 < H_{ex}$, for negative field variations; here $H_{en}$ and $H_{ex}$ are the threshold fields defined by Clem [11] for vortex entry and exit, respectively. During the field sweep, the fluxons that cannot follow the magnetic-field variations are accumulating near the sample surface. As soon as the field sweep is stopped, a diffusive motion of fluxons sets in; the process ends when the flux density reaches the appropriate value for the critical state. This process may account for the exponential decay of the SH signal in the time scale of seconds. At the same time, motion of fluxons through the surface barrier takes place and it may be responsible for the logarithmic decay of the SH signal. The latter hypothesis is corroborated by the fact that the logarithmic-decay rates of the SH signal are different for magnetic fields reached at increasing and decreasing values.

## References

[1] S. T. Weir *et al.*, Phys. Rev. **B 43** (1991) 3034.
[2] M. Konczykowski *et al.*, Phys. Rev. **B 43** (1991) 13707.
[3] N. Chikumoto *et al.*, Phys. Rev. Lett. **69** (1992) 1260.
[4] M. Pissas *et al.*, J. Supercond. **14** (2001) 615.
[5] A. Agliolo Gallitto *et al.*, cond-mat/0306467, Physica **C**, in press.
[6] I. Ciccarello *et al.*, Physica **C 159** (1989) 769.
[7] L. Burlachkov, Pys. Rev. **B 47** (1993) 8056.
[8] D. Zola, *et al.*, Int. J. Mod. Phys. **14** (2000) 25.
[9] E. H. Brandt, Rep. Prog. Phys. **58** (1995) 1465.
[10] Y. Takano *et al.*, Appl. Phys. Lett. **78** (2001) 2914.
[11] J. R. Clem, *Proc. LTP13$^{th}$ Conference,* K. D. Timmerhaus *et al.* ed.s (Plenum, New York, 1974), vol.3, p.102.